\documentclass[conference]{IEEEtran}
\IEEEoverridecommandlockouts
\usepackage{cite}
\usepackage{amsmath,amssymb,amsfonts}
\usepackage{cases}
\usepackage{algorithmic}
\usepackage{graphicx}
\usepackage{textcomp}
\usepackage{xcolor}

\DeclareMathOperator{\R}{\mathbb{R}} 
\DeclareMathOperator{\Z}{\mathbb{Z}} 

\newcommand{\p}{\mathbb{P}} 
\newcommand{\E}{\mathbb{E}} 

\def\cA{{\mathcal A}}

\def\cF{{\mathcal F}}

\newcommand{\Ntime}{\mathbb{T}^N}

\newcommand{\note}[1]{{\color{blue}{\bf [#1]}}}
\newcommand{\cut}[1]{}

\newtheorem{theorem}{Theorem}[section]

\newtheorem{example}[theorem]{Example}

\def\BibTeX{{\rm B\kern-.05em{\sc i\kern-.025em b}\kern-.08em
    T\kern-.1667em\lower.7ex\hbox{E}\kern-.125emX}}
\begin{document}

\title{On the Accuracy of Deterministic Models for Viral Spread on Networks\\
\thanks{The work of Anirudh Sridhar is supported by Army Research Office Grant W911NF-20-1-0204, the C3.ai Digital Transformation Institute, and the National Science Foundation under grants IIS-2026982 and DMS-1811724. The work of Soummya Kar is partially supported by the National Science Foundation under grant CNS-1837607.}
}

\author{\IEEEauthorblockN{Anirudh Sridhar}
\IEEEauthorblockA{\textit{Department of Electrical and Computer Engineering} \\
\textit{Princeton University}\\
Princeton, NJ \\
anirudhs@princeton.edu}
\and
\IEEEauthorblockN{Soummya Kar}
\IEEEauthorblockA{\textit{Department of Electrical and Computer Engineering} \\
\textit{Carnegie Mellon University}\\
Pittsburgh, PA \\
soummyak@andrew.cmu.edu}
}

\maketitle

\begin{abstract}
We consider the emergent behavior of viral spread when agents in a large population interact with each other over a contact network. When the number of agents is large and the contact network is a complete graph, it is well known that the population behavior -- that is, the fraction of susceptible, infected and recovered agents -- converges to the solution of an ordinary differential equation (ODE) known as the classical SIR model as the population size approaches infinity. In contrast, we study interactions over contact networks with generic topologies and derive conditions under which the population behavior concentrates around either the classic SIR model or other deterministic models. Specifically, we show that when most vertex degrees in the contact network are sufficiently large, the population behavior concentrates around an ODE known as the network SIR model. We then study the short and intermediate-term evolution of the network SIR model and show that if the contact network has an expander-type property or the initial set of infections is well-mixed in the population, the network SIR model reduces to the classical SIR model. To complement these results, we illustrate through simulations that the two models can yield drastically different predictions, hence use of the classical SIR model can be misleading in certain cases. 
\end{abstract}

\begin{IEEEkeywords}
epidemic models, mean-field approximation, network-based interactions, SIR models
\end{IEEEkeywords}

\section{Introduction}

The emergence of infectious diseases has led to devastating public health and economic crises worldwide. It is therefore of the utmost importance to develop accurate mathematical models of viral spread in order to track outbreaks and devise effective mitigation strategies. In the basic model, known as the {\it stochastic susceptible-infected-recovered (SIR) process}, there is a large population of agents who are initially either susceptible, infected or recovered. As agents interact with each other over time, susceptible agents may become infected if they interact with infected agents, and infected agents eventually recover (see, e.g., \cite[Chapter 9.3]{brauer2012mathematical}). When the number of agents is large, studying the individual-level evolution of the epidemic is typically intractable from both theoretical and computational perspectives. To get around this issue, researchers often utilize a so-called {\it mean-field approximation} which is a an ordinary differential equation (ODE) tracking the fraction of susceptible, infected and recovered agents in the population. This ODE, which we call the {\it classical SIR model} \cite{SIR_og}, has been widely used for nearly a century to understand the spread of an epidemic. 

A common criticism of the classical SIR model is that it assumes an all-to-all interaction structure: the probability that two given agents interact is the same for {\it any} pair of agents. In reality, agent interactions are constrained by a {\it contact network}, which is known to have a significant effect on the spread of a virus. To address this gap, researchers have proposed a generalization of the classical SIR model that incorporates the contact network, which we shall call the {\it network SIR model} \cite{lajmanovich1976, wang_sis}. This model has received significant attention over the past few decades, leading to exact computations of key epidemiological properties for general interaction structures as well as applications to parameter estimation and control \cite{mei_network_epidemics, khanafer2016, mieghem2009, pare_analysis_control, preciado_control, mai_suppressing}. Despite the multitude of papers on the network SIR model, fundamental questions remain open. For instance, it is unclear whether the network SIR model can reasonably approximate the stochastic SIR process. Furthermore, since the network SIR model is challenging to simulate when the size of the population is large, little is known about the short and intermediate-term evolution of the model. 

The goal of our paper is to address these gaps. First, we make use of a recent general result of Sridhar and Kar \cite{sridhar2021meanfield} to show that if most vertices in the underlying contact network have a large enough degree, then the network SIR model correctly predicts the population-level behavior of the stochastic SIR model. Typically, to prove that a stochastic process concentrates around a deterministic counterpart, one would appeal to a law of large numbers that holds when the size of the population grows to infinity. The main difficulty is that the stochastic and network SIR models both depend heavily on the {\it finite} interaction structure between agents, hence no such law of large numbers exists in this case. This makes the mathematical analysis quite challenging, which is a key reason why the network SIR model was previously not rigorously justified. 

Next, we study the short and intermediate-term behavior of the network SIR model. We find that the network SIR model is {\it equivalent} to the classical SIR model if the contact network satisfies an {\it expander-type} property or if the initial infections are well-mixed within the population. This is somewhat surprising since the classical SIR model -- which assumes a simplistic, homogenous interaction structure -- can yield accurate predictions for arbitrary interaction structures. These results are established through novel connections to notions of {\it consensus} from distributed control. While such ideas are also considered in the general framework studied in \cite{sridhar2021meanfield}, our analysis departs from their work by specializing to the network SIR model and by removing restrictive assumptions on the interaction structure used in \cite{sridhar2021meanfield}. 

Finally, through simulations, we show that the classical and network SIR can yield drastically different predictions for the spread of an epidemic. For instance, it is well-known that if an epidemic emerges, the classical SIR model predicts a single epidemic peak, after which the infected population dies out. We simulated the network SIR model for spatially-structured contact networks and found that when the average degree in the network is large, we see the same {\it qualitative} behavior as the classical SIR model (e.g., the emergence of a single epidemic peak). For networks with smaller average degree, there is an initial decline in infections followed by a resurgence at a later point once the epidemic has spread to other parts of the network. This shows in particular that the network SIR model may be able to predict the emergence of multiple waves of infection observed in reality \cite{multiple_peaks}. The classical SIR model, on the other hand, does not capture this phenomenon.

The structure of the paper is as follows. In Section \ref{sec:notation}, we define relevant notation. Section \ref{sec:models} reviews the literature on the stochastic SIR process, the classical SIR model and the network SIR model. In Section \ref{sec:approx}, we leverage the results from \cite{sridhar2021meanfield} to study the error between the stochastic SIR process and the network SIR model. In Section \ref{sec:equivalence} we identify cases where the network SIR model reduces to the classical SIR model. Section \ref{sec:simulations} includes our simulations. Finally, we conclude in Section \ref{sec:conclusion}.

\section{Notation}
\label{sec:notation}

Denote $\R, \Z$ and $\Z_{\ge 0}$ to be the set of reals, integers and non-negative integers, respectively. For a positive integer $K$, denote $[K] : = \{ 1, \ldots, K \}$. For a vector $v \in \R^k$, the $p$-norm is $\| v \|_p : = ( \sum_{i = 1}^k |v_i |^p )^{1/p}$. The infinity norm is $\| v \|_\infty : = \max_{i \in [k] } | v_i |$. For a matrix $W \in \R^{k \times m}$, the {\it Frobenius norm} of $W$ is $\| W \|_F : = ( \sum_{i = 1}^k \sum_{j = 1}^m w_{ij}^2 )^{1/2}$.

\section{Epidemic models}
\label{sec:models}

\subsection{The stochastic SIR process}
\label{subsec:stoch}

We begin by reviewing the rigorous construction of the stochastic SIR process. The parameters of the virus are given by $\beta, \gamma \in [0,1]$, which denote the probabilities of transmission and recovery, respectively. Suppose we have a population of $N$ agents, indexed by elements of $[N]$, who are always in one of three states: $(S)$, infected $(I)$ or recovered $(R)$. We also assume the existence of an undirected contact network $G$ connecting the $N$ agents, so that if $(i,j) \in E(G)$, it is possible for $i$ and $j$ to interact with each other. It is assumed that each agent has an independent {\it Poisson clock}, which is a rate-1 Poisson process on the non-negative reals. When an agent's clock ``rings" (equivalently, the agent's associated Poisson process jumps), the agent updates their state. If the updating agent is susceptible, they interact with a randomly chosen agent in their neighborhood. If the latter agent is infected, they transmit the virus to the updating agent (i.e., the state changes from $S$ to $I$) with probability $\beta$. If, on the other hand, the updating agent is infected, they recover (i.e., the state changes from $I$ to $R$) with probability $\gamma$.\footnote{An equivalent and equally common assumption is that the waiting time between interactions and recovery are independent exponential random variables.}
 
Due to the possibly complex interactions between agents, it is typically quite challenging to simulate and analyze the time-evolution of the stochastic SIR process when $N$ is large. Most of the theoretical literature on this process has studied time-asymptotic properties such as steady-state behavior using branching process methods and non-rigorous techniques from theoretical physics (see \cite[Section 4]{ps2015} and references therein). A key insight from this literature is that the emergence and final size of an epidemic depends not only on the properties of the virus, but also on the structure of the contact network. 

\subsection{Classical mean-field models}

A special case where the stochastic SIR process admits a tractable analysis is the so-called {\it homogenous mixing} case, in which the interactions between any pair of agents are equally likely. This implies that the contact network is a complete graph (i.e., all-to-all links) and that when a susceptible agent is chosen to update, they interact with another agent chosen {\it uniformly at random} from the entire population. In the limit of large populations, the resulting dynamics can be approximated by the following ODE: 
\begin{equation}
\label{eq:classical_mf}
\begin{cases}
\dot{x}^S(t)  = - \beta x^S(t) x^I(t) & \\
\dot{x}^I(t)  = \beta x^S(t) x^I(t) - \gamma x^I(t) & \\
\dot{x}^R(t)  = \gamma x^I(t), &
\end{cases}
\end{equation}
where $x^S(t), x^I(t), x^R(t)$ represent the fraction of susceptible, infected and recovered agents in the population, respectively. The model \eqref{eq:classical_mf}, which we call the {\it classical SIR model}, can be justified as follows. Suppose that at time $t$, a single agent -- labeled by $i$ -- is chosen to update. If $i$ is susceptible, which occurs with probability $x^S(t)$, then they interact with another agent -- labeled by $j$ -- chosen uniformly at random from the population. If $j$ is infected, which occurs with probability $x^I(t)$, $i$ changes state from $S$ to $I$ with probability $\beta$, else $i$ remains in state $S$. Putting everything together, the probability of a single agent changing from state $S$ to $I$ at time $t$ is $\beta x^S(t) x^I(t)$. In the limit of large populations, this change is infinitesimal with respect to the population, which explains the first equation in \eqref{eq:classical_mf}. The observation that if $i$ is infected -- which occurs with probability $x^I(t)$ -- they recover with probability $\gamma$ explains the second and third equations in \eqref{eq:classical_mf}. 

The ODE \eqref{eq:classical_mf} is relatively simple, easy to simulate and provides population-level information about the spread of a virus {\it without} needing to keep track of the states of individual agents. The original model of this type was proposed by Kermack and McKendrick in 1927 \cite{SIR_og}. Various generalizations and extensions have been studied over the years, including additional state transitions (e.g., SEIR and SIS models, see \cite[Chapter 9.5]{brauer2012mathematical}) and multi-population models \cite{Watts2005,hanski1999metapopulation}. 

The classical SIR model is also rigorously justified from a mathematical point of view. Given an instantiation of the stochastic SIR process on an $N$-agent population, define the {\it population state}
$$
Y_{av}^N(t) : = ( Y_{av}^{N,S}(t), Y_{av}^{N,I}(t), Y_{av}^{N,R}(t) )
$$
where $Y_{av}^{N,S}(t)$ is the fraction of agents in state $S$, with similar interpretations for $Y_{av}^{N,I}(t)$ and $Y_{av}^{N,R}(t)$. Due to well-known results of Kurtz \cite{Kurtz1976,kurtz}, if $Y_{av}^N(0) \to x(0)$ as $N \to \infty$, then 
$$
\p \left( \lim\limits_{N \to \infty} \sup_{0 \le t \le T} \| Y_{av}^N(t) - x(t) \|_\infty = 0 \right) = 1.
$$

\subsection{A mean-field model with non-homogenous interactions}

A significant drawback of the classical SIR model is that the homogenous mixing assumption is quite unrealistic, since agents may be more or less likely to interact with others based on their physical locations and friend circles. To address this gap, Lajmanovich and Yorke \cite{lajmanovich1976} and later Wang et al \cite{wang_sis} proposed a generalization of the classical model accounting for {\it non-homogenous} interactions between agents. The model is defined as follows. Suppose we have a finite population of $N$ agents. For each agent $i$, define the variable
$$
y_i^N(t) : = \{ y_i^{N,S}(t), y_i^{N,I}(t), y_i^{N,R}(t) \},
$$
where $y_i^{N,S}(t)$ represents the probability that agent $i$ is in state $S$ at time $t$, with similar interpretations for $y_i^{N,I}(t)$ and $y_i^{N,R}(t)$. Next, define the {\it interaction matrix} $W \in \R^{N \times N}$, where the entry $w_{ij}$ represents the probability that $i$ interacts with $j$. The sparsity of $W$ conforms to the structure of the contact network $G$ in the following sense: $w_{ij} > 0$ if and only if $(i,j) \in E(G)$. We further note that, by the law of total probability, $\sum_{j = 1}^N w_{ij} = 1$ for each $i \in [N]$ so $W$ is a row-stochastic matrix. The distribution of states within $i$'s neighborhood is then given by 
$$
\overline{y}_i^N(t) : = \sum\limits_{j = 1}^N w_{ij} y_j^N(t),
$$
so that, in particular, $\overline{y}_i^{N,I}(t)$ is the probability that agent $i$ interacts with an infected neighbor at time $t$. The {\it network SIR model} is the following $3N$-dimensional system of ODEs:

\begin{equation}
\label{eq:network_mf}
\begin{cases}
\dot{y}_i^{N,S}(t)  = - \beta y_i^{N,S}(t) \overline{y}_i^{N,I}(t) & \\
\dot{y}_i^{N,I}(t)  = \beta y_i^{N,S}(t) \overline{y}_i^{N,I}(t) - \gamma y_i^{N,I}(t) & \\
\dot{y}_i^{N,R}(t)  = \gamma y_i^{N,I}(t).  \qquad i \in [N] &
\end{cases}
\end{equation}

The model \eqref{eq:network_mf} can be justified in a similar manner to \eqref{eq:classical_mf}. Suppose that agent $i$ is chosen to update at time $t$. If $i$ is infected -- which occurs with probability $y_i^{N,I}(t)$ -- then they recover with probability $\gamma$, thus justifying the $\gamma y_i^{N,I}(t)$ in the second and third equations of \eqref{eq:network_mf}. If $i$ is susceptible -- which occurs with probability $y_i^{N,S}(t)$ -- and they interact with an infected neighbor -- which occurs with probability $\overline{y}_i^{N,I}(t)$ -- then agent $i$ changes from $S$ to $I$ with probability $\beta$. If agent states are independent, the combined probability is $\beta y_i^{N,S}(t) \overline{y}_i^{N,I}(t)$, which justifies the term in the first and second equations of \eqref{eq:network_mf}. A flaw in this explanation is that agent states are in general {\it not independent} since neighboring agents may interact with each other. However, it is expected that if agent neighborhoods are sufficiently large, the $y_i^N(t)$'s may be {\it approximately} independent, in which case the behavior of \eqref{eq:network_mf} may correctly align with the stochastic SIR process.

Like the stochastic SIR process, the network SIR model is high-dimensional and therefore challenging to simulate when $N$ is large. However, a theoretical analysis of \eqref{eq:network_mf} is often more tractable than its stochastic counterpart because it is deterministic and one can apply classical ODE methods in a relatively straightforward manner to establish stability and rate of convergence to equilibria \cite{lajmanovich1976, mei_network_epidemics, khanafer2016, mieghem2009}. Notably, this permits an analysis of {\it general} contact networks, whereas work on the stochastic SIR process typically assumes the contact network is highly structured (e.g., drawn from a random graph family such as the configuration model) \cite{Newman_2002, massoulie_epidemics, graphon_epidemics}. The network SIR model also offers a tractable baseline model to study the estimation of propagation dynamics and epidemic control on networks \cite{pare_analysis_control, preciado_control, mai_suppressing}.

Despite the multitude of papers on the network SIR model, there is little work comparing the predictions of the network SIR model and the stochastic SIR process. Van Miegham, Omic and Kooij \cite{mieghem2009} as well as Cator and Van Miegham \cite{cator_approx} proved that in a network SIS model, $y_i^{N,I}(t)$ is an {\it upper bound} for the probability that $i$ is infected at time $t$. Van Miegham and van de Bovenkamp additionally derived an expression for the error between \eqref{eq:network_mf} and the {\it expected} behavior of the corresponding stochastic SIR process in terms of the covariances between the $y_i^N(t)$'s. They estimate the error analytically for the complete graph and star graph and empirically for Erd\H{o}s-R\'{e}nyi graphs. They further conjecture that the network SIR model is accurate when the average degree is large \cite{mieghem_approx}. \cut{Additionally, though time-asymptotic properties of the network SIR model, such as convergence to equilibria, have been mostly determined \note{cite}, a clear picture of the short-term and intermediate-term dynamics is lacking.}

\section{Deterministic approximation of the stochastic SIR process}
\label{sec:approx}

In this section, we leverage the recent results of Sridhar and Kar \cite{sridhar2021meanfield} to show that in the case of non-homogenous interactions, the population state $Y_{av}^N(t)$ can be well-approximated by the process 
$$
y_{av}^N(t) : = \frac{1}{N} \sum\limits_{i = 1}^N y_i^N(t)
$$
derived from the network SIR model. We first review the general class of stochastic processes considered in \cite{sridhar2021meanfield}. The set of agent states is denoted by a finite set $\cA$, and initially, each of the $N$ agents has a state in $\cA$. For each distinct $a,b \in \cA$, define the {\it rate function} 
$$
\rho^{ab}: \left \{ z \in \R^{ | \cA|} : \sum\limits_{a \in \cA} z_a = 1 \land  z_a \ge 0, \forall a \in \cA \right \} \to [0,1].
$$
The rate functions are assumed to be Lipschitz. Next, define the vectorized state of agent $i$ to be $Y_i^N(t) : = \{Y_i^{N,a}(t) \}_{a \in \cA}$, where $Y_i^{N,a}(t) = 1$ if agent $i$ has state $a$ at time $t$, else it is 0. We may also define $\overline{Y}_i^N(t) : = \{ \overline{Y}_i^{N,a}(t) \}_{a \in \cA}$, where
$$
\overline{Y}_i^{N,a}(t) : = \sum\limits_{j= 1}^N w_{ij} Y_j^{N,a}(t).
$$
For each transition time $t$ in the set 
$$
\Ntime : = \left \{ 0 , \frac{1}{N}, \frac{2}{N} , \ldots \right \} = \left \{ \frac{k}{N} : k \in \Z_{\ge 0} \right \},
$$
a single agent is chosen uniformly at random to update their state.\footnote{This can be viewed as a discretization of the Poisson clock model discussed in Section \ref{subsec:stoch}.} If the updating agent has state $a$, they change to state $b$ with probability $\rho^{a b} ( \overline{Y}_i^N(t))$. The SIR process is a special case of the stochastic dynamics described above, with $\cA = \{S, I, R \}$ and rate functions given by $\rho^{SI}(z) = \beta z^I$ and $\rho^{IR}(z) = \gamma$, with the other rate functions being zero. Since $\rho^{SI}$ is linear and $\rho^{IR}$ is constant, the Lipschitz assumption on the rate functions is satisfied.

The following result is a consequence of specializing a main result of Sridhar and Kar \cite[Theorem 3.2]{sridhar2021meanfield} to the SIR model. 

\begin{theorem}
\label{thm:concentration}
Suppose that $W$ is non-negative and doubly-stochastic, and that $\beta, \gamma < 1$. Fix $\epsilon > 0$ and a time horizon $T \ge 0$. There exists a constant $L = L(\beta, \gamma)$ such that if $N \ge 4 e^{LT}/\epsilon$ and 
\begin{equation}
\label{eq:frobenius_norm}
\frac{1}{N} \| W \|_F^2 : = \frac{1}{N} \sum\limits_{i = 1}^N \sum\limits_{j = 1}^N w_{ij}^2 \le \frac{\epsilon^2}{8T e^{2LT}},
\end{equation}
and if $\hat{Y}_{av}^N(t)$ is the continuous-time version of $Y_{av}^N(t)$ formed by linear interpolation between successive values in $\Ntime$, then 
\begin{equation}
\label{eq:prob_bound}
\p \left( \max\limits_{0 \le t \le T} \| \hat{Y}_{av}^N(t) - y_{av}^N(t) \|_\infty > \epsilon \right) \le c_1 e^{- c_2 N \epsilon^2},
\end{equation}
where $c_1 = c_1 (T, \epsilon, \beta , \gamma)$ and $c_2 = c_2 (T, \beta, \gamma)$. 
\end{theorem}

Equation \eqref{eq:prob_bound} shows that when $\epsilon, T$ are fixed and $N$ is large, $y_{av}(t)$ is a good approximation for $Y_{av}^N(t)$. Moreover, Theorem \ref{thm:concentration} establishes a large-deviations-type probability upper bound, which is similar to large-deviations results for the homogenous mixing case \cite{BenaimWeibull, sandholm_staudigl_2018}.

The key technical assumption that enables Theorem \ref{thm:concentration} is \eqref{eq:frobenius_norm}. At a high level, \eqref{eq:frobenius_norm} ensures that the underlying contact network is not too sparse; we illustrate this concretely through the following example. 
\begin{example}
\label{ex:contact_network}
Let $G$ be the underlying contact network, and let $d_i$ be the degree of agent $i$. Construct $W$ so that $w_{ij} = 1/d_i$ if $(i,j) \in E(G)$, else $w_{ij} = 0$. Then 
$$
\frac{1}{N} \| W \|_F^2 = \frac{1}{N} \sum\limits_{i = 1}^N \frac{1}{d_i}. 
$$
For this quantity to be sufficiently small, all of the $d_i$'s except for a small fraction of vertices must be sufficiently large.  
\end{example}
We remark that $W$ constructed in Example \ref{ex:contact_network} is not doubly-stochastic in general. However, the doubly-stochastic condition is used in \cite{sridhar2021meanfield} mainly for convenience as it simplifies much of the analysis, so we expect that this condition can be relaxed.

\section{Equivalence of the classical and network mean-field SIR models}
\label{sec:equivalence}

While Theorem \ref{thm:concentration} provides an important first step in understanding the behavior of the stochastic SIR process with non-homogenous interactions, an important remaining task is to study the behavior of $y_{av}^N(t)$. Although this is a challenging task in general as it entails the analysis of the $N$-dimensional ODE \eqref{eq:network_mf}, we show in the following section that $y_{av}^N(t)$ reduces to the classical mean-field SIR model in many cases of interest. 

The key insight behind this reduction is that if the collection $\{\overline{y}_i^N(t) \}_{i \in [N]}$ is at a {\it consensus} -- that is, $\overline{y}_i^N(t) = y_{av}^N(t)$ for all $i$ -- then $y_{av}^N(t)$ is {\it exactly} equal to a solution of the classical mean-field ODE \eqref{eq:classical_mf}. Indeed, if we make the substitution $\overline{y}_i^N(t) = y_{av}^N(t)$ in \eqref{eq:network_mf}, we have
\begin{equation}
\begin{cases}
\dot{y}_i^{N,S}(t)  = - \beta y_i^{N,S}(t) y_{av}^{N,I}(t) & \\
\dot{y}_i^{N,I}(t) = \beta y_i^{N,S}(t) y_{av}^{N,I}(t) - \gamma y_i^{N,I}(t) & \\
\dot{y}_i^{N,R}(t) = \gamma y_i^{N,I}(t). &
\end{cases}
\end{equation}
Averaging over $i$, we see that $y_{av}(t)$ satisfies
\begin{equation}
\begin{cases}
\dot{y}_{av}^{N,S}(t) = - \beta y_{av}^{N,S}(t) y_{av}^{N,I}(t) & \\
\dot{y}_{av}^{N,I}(t) = \beta y_{av}^{N,S}(t) y_{av}^{N,I}(t) - \gamma y_{av}^{N,I}(t) & \\
\dot{y}_{av}^{N,R}(t) = \gamma y_{av}^{N,I}(t), &
\end{cases}
\end{equation}
so $y_{av}^N(t)$ is a solution to the classical mean-field ODE \eqref{eq:classical_mf}. Using perturbation arguments, it can be shown that if the $\overline{y}_i^N(t)$'s are {\it close} to a consensus, then $y_{av}^N(t)$ is close to a solution of \eqref{eq:classical_mf}. We show that this is possible under certain generic assumptions on the interaction matrix as well as the locations of initial infections. Sridhar and Kar \cite{sridhar2021meanfield} investigated these cases under the restrictive assumption that $W$ is doubly-stochastic. In the context of epidemic modeling, requiring $W$ to be doubly stochastic is quite unrealistic, as the constraint that all {\it column} sums must be one enforces a strange coupling across the interaction probabilities of all agents. In this work, we therefore present an analysis specialized to the network SIR model that holds for the much broader class of {\it row-stochastic} matrices. 

\noindent {\bf Properties of the interaction matrix.} In many cases of interest, the structure of $W$ may guarantee that the $\overline{y}_i^N(t)$'s are, on average, close to a consensus. To capture this idea formally, suppose that $W \in \R^{N \times N}$ and $\mathbf{1} \in \R^N$ is the vector of all ones. Define
$$
\lambda(W) : = \sup\limits_{x : \| x \|_2 = 1, \langle \mathbf{1}, x \rangle = 0 } \| Wx \|_2.
$$
Since $W$ is row-stochastic, $\mathbf{1}$ is an eigenvector with eigenvalue 1. In the special case where $W$ is symmetric, all other eigenvectors are orthogonal to $\mathbf{1}$, hence $\lambda(W)$ is achieved by an eigenvector of $W$ corresponding to the second-largest eigenvalue in magnitude.  

At a high level, $\lambda(W)$ controls the deviation between $y_{av}(t)$ and the solution to \eqref{eq:classical_mf} with the same initial conditions. To see why this is the case, write 
$$
\dot{y}_{av}^{N,S}(t) = - \beta y_{av}^{N,S}(t) y_{av}^{N,I}(t) - p(t),
$$,
where $p(t)$ is a perturbation given explicitly by
$$
p(t) : = \frac{1}{N} \sum\limits_{i = 1}^N \beta y_i^{N,S}(t) ( \overline{y}_i^{N,I}(t) - y_{av}^{N,I}(t) ).
$$
Define $\mathbf{y}^{N,S}(t) : = \{ y_i^{N,S}(t) \}_{i = 1}^N$ and $\overline{\mathbf{y}}^{N,S}(t) : = \{ \overline{y}_i^{N,S}(t) \}_{i = 1}^N$, and note that $\overline{\mathbf{y}}^{N,S}(t) = W \mathbf{y}^{N,S}(t)$. We can then bound
\begin{align}
\label{eq:a}
|p(t)|^2 & \le \frac{1}{N} \sum\limits_{i = 1}^N ( \beta y_i^{N,S}(t) )^2 ( \overline{y}_i^{N,I}(t) - y_{av}^{N,I}(t) )^2 \\
\label{eq:b}
& \le \frac{1}{N} \| \overline{\mathbf{y}}^{N,I}(t) - y_{av}^{N,I}(t) \mathbf{1} \|_2^2 \\
& = \frac{1}{N} \| W ( \mathbf{y}^{N,I}(t) - y_{av}^{N,I}(t) \mathbf{1} ) \|_2^2 \nonumber \\
\label{eq:c}
& \le \frac{\lambda(W)^2}{N} \| \mathbf{y}^{N,I}(t) - y_{av}^{N,I}(t) \mathbf{1} \|_2^2 \\
\label{eq:d}
& \le \lambda(W)^2.
\end{align}
Above, \eqref{eq:a} is due to Jensen's inequality, \eqref{eq:b} is due to $| \beta y_i^{N,S}(t) | \le 1$ and the definition of the $\ell_2$ norm, \eqref{eq:c} follows since $\langle \mathbf{1}, \mathbf{y}^{N,I}(t) - y_{av}^{N,I}(t) \mathbf{1} \rangle = 0$ and \eqref{eq:d} follows from $| y_i^{N,I}(t) - y_{av}^{N,I}(t) | \le 1$ for all $i$. Hence we expect that if $\lambda(W)$ is small, $y_{av}^N(t)$ will resemble a solution to \eqref{eq:classical_mf}. This leads to the following result. 

\begin{theorem}
\label{thm:rapidly_mixing}
Let $\{W_n \}_n$ be a sequence of row-stochastic matrices such that $W_n \in \R^{n \times n}$ and 
\begin{equation}
\label{eq:rapidly_mixing}
\lim\limits_{n \to \infty} \lambda(W_n) = 0.
\end{equation}
Then for any time horizon $T \ge 0$, 
$$
\lim\limits_{n \to \infty} \sup\limits_{0 \le t \le T} \| y_{av}^n(t) - x(t) \|_\infty = 0,
$$
where $x(t)$ is a solution to \eqref{eq:classical_mf} with $x(0) = y_{av}^n(0)$. 
\end{theorem}

The following example shows that when the underlying contact network is an Erd\H{o}s-R\'{e}nyi graph, \eqref{eq:rapidly_mixing} can be satisfied. 

\begin{example}
\label{ex:ER}
We say that $G \sim G(n,p)$ if $G$ is a graph on $[n]$ and for each pair $(i,j) \in [n]^2$ such that $i \neq j$, the edge $(i,j)$ is included in $G$ with probability $p$, independently across all pairs of vertices. Let $d_i$ be the degree of vertex $i$, and let the entries of $W$ be given by $w_{ij} = 1/d_i$ if $(i,j) \in E(G)$. With high probability, if $p$ is asymptotically larger than $O (\log^4 (n) / n)$, 1 is an eigenvalue of $W$ with multiplicity 1, and the magnitude of all other eigenvalues are at most $O ( 1/ \sqrt{np})$ \cite{Chung}. Since the degrees of all vertices are tightly concentrated around $np$ for the regime of $p$ we consider, $W$ is {\it nearly} symmetric. We therefore expect that $\lambda(W)$ is {\it close} to the second-largest eigenvalue in magnitude of $W$, and \eqref{eq:rapidly_mixing} follows. 
\end{example}

\noindent {\bf Distribution of initial infections.} In general,  $\lambda(W)$ may {\it not} close to 0 so we do not expect that $y_{av}^N(t)$ can be well-approximated by a solution to \eqref{eq:classical_mf}; this is illustrated empirically in Section \ref{sec:simulations}, Figures \ref{fig:nn_localized} and \ref{fig:nn_deterministic}. However, the dynamics \eqref{eq:network_mf} have a convenient {\it consensus-stable} property: if the $\overline{y}_i^N(t)$'s are close to a consensus initially, then they {\it remain} close to a consensus for a long time. To see why this property holds, we first compute the following derivative: 
\begin{align*}
&\frac{d}{dt} ( \overline{y}_i^{N,S}(t) - y_{av}^{N,S}(t) ) \\
& = - \beta \left( \sum\limits_{j = 1}^N w_{ij} y_j^{N,S}(t) \overline{y}_j^{N,I}(t) - y_{av}^{N,S}(t) y_{av}^{N,I}(t)  \right) + p(t) \\
& = - \beta \left( \sum\limits_{j = 1}^N w_{ij} y_j^{N,S}(t) ( \overline{y}_j^{N,I}(t) - y_{av}^{N,I}(t) ) \right. \\
& \left. \vphantom{\sum\limits_{j= 1}^N} \hspace{1.5cm} +  (\overline{y}_i^{N,S}(t) - y_{av}^{N,S}(t) ) y_{av}^{N,I}(t)  \right) + p(t).
\end{align*}
Since $y_j^{N,S}(t) , y_{av}^{N,I}(t) \in [0,1]$, $p(t) \le \max_{i \in [N]} \| \overline{y}_i^N(t) - y_{av}^N(t) \|_\infty$ in light of \eqref{eq:b} and $\sum_{j= 1}^N w_{ij} = 1$, we have the bound
\begin{equation}
\label{eq:baryi_derivative_bound}
\left | \frac{d}{dt} ( \overline{y}_i^{N,S}(t) - y_{av}^{N,S}(t) ) \right | \le 4 \max\limits_{i \in [N]} \| \overline{y}_i^N(t) - y_{av}^N(t) \|_\infty.
\end{equation}
Through a similar analysis, it can be shown that the bound in \eqref{eq:baryi_derivative_bound} also holds for the states $I$ and $R$. Taking a maximum over $i \in [N]$ and $a \in \{S, I, R \}$ shows that the following inequality holds for almost all $t \ge 0$:\footnote{The derivative of a maximum of finitely many differentiable functions may be non-differentiable when two of the functions intersect at a single point, which can happen only on a set of Lebesgue measure zero.}
$$
\frac{d}{dt} \left( \max\limits_{i \in [N]} \| \overline{y}_i^N(t) - y_{av}^N(t) \|_\infty  \right) \le 4 \max\limits_{i \in [N]} \| \overline{y}_i^N(t) - y_{av}^N(t) \|_\infty.
$$
The differential form of Gr\"{o}nwall's inequality implies that
\begin{equation*}
\label{eq:infty_norm_bound}
\max\limits_{i \in [N]} | \overline{y}_i^N(t) - y_{av}^N(t) \|_\infty \le \left( \max\limits_{i \in [N] } \| \overline{y}_i^N(0) - y_{av}^N(0) \|_\infty \right) e^{4t}. 
\end{equation*}
Noting that the bound in the display above also holds for the perturbation $p(t)$, it follows that $\sup_{0 \le t \le T} p(t)$ can be made arbitrarily small as long as $\max_{i \in [N]} \| \overline{y}_i^N(0) - y_{av}^N(0) \|$ is small as well. This leads to the following result. 

\begin{theorem}
\label{thm:consensus_stable}
Suppose that $x$ is a solution to \eqref{eq:classical_mf} with $x(0) = y_{av}^N(0)$. For every $\epsilon > 0$, there exists $\delta = \delta(T, \epsilon, \beta, \gamma)$ such that if 
$$
\max\limits_{i \in [N] } \| \overline{y}_i^N(0) - y_{av}^N(0) \|_\infty \le \delta,
$$
then
\begin{equation*}
\label{eq:consensus_stable}
\sup\limits_{0 \le t \le T} \| y_{av}^N(t) - x(t) \|_\infty \le \epsilon.
\end{equation*}
\end{theorem}

A natural scenario in which we may expect $\| \overline{y}_i^N(0) - y_{av}^N(0) \|_\infty$ to be small is when the set of initial infections is randomly interspersed throughout the population. This is concretely illustrated in the following example. 

\begin{example}
\label{ex:random_initial_conditions}
Let the contact network be a $d$-regular graph, with $w_{ij} = 1/d$ if $(i,j)$ is an edge in the network, else $w_{ij} = 0$. Suppose that, independently at random across all agents, each agent is infected with probability $q$ and susceptible with probability $1 - q$. Hoeffding's inequality implies
\begin{align*}
\p ( \| y_{av}^{N}(0)  - (1-q,q,0) \|_\infty > \epsilon ) & \le 2 e^{ - 2 N \epsilon^2} \\
\p ( \| \overline{y}_i^{N}(0) - (1-q,q,0) \|_\infty > \epsilon ) & \le 2 e^{ - 2 d \epsilon^2}, \qquad i \in [N].
\end{align*}
A union bound then implies that
$$
\p \left( \max\limits_{i \in [N] } \| \overline{y}_i^N(0) - y_{av}^N(0) \|_\infty > \epsilon \right) \le 2N e^{- \frac{1}{2} d \epsilon^2 }.
$$
In particular, if $d/ \log N$ is sufficiently large, the right hand side of \eqref{eq:consensus_stable} can be made arbitrarily small for large $N$. 
\end{example}

\section{Simulations}
\label{sec:simulations}

In this section, we provide several simulations of the classical mean-field SIR model \eqref{eq:classical_mf}, the network mean-field SIR model \eqref{eq:network_mf} and the stochastic SIR model to support our theoretical results. In all our simulations, we set $\beta = 0.8$ and $\gamma = 0.3$; we found that these parameters demonstrated the key qualitative aspects of epidemic spread such as a peak in infections and a subsequent exponential decay. In Section \ref{subsec:concentration_sim}, we empirically support our results in Sections \ref{sec:approx} and \ref{sec:equivalence} which collectively determine conditions under which the population process $Y_{av}^N(t)$ concentrates around the classical mean-field SIR model. In Section \ref{subsec:deviation_sim}, we highlight cases where $Y_{av}^N(t)$ concentrates around the {\it network mean-field SIR model} as opposed to the classical model. In particular, we show how the location of initial conditions and the structure of the underlying contact network has a significant effect on the emergence and ultimate size of an epidemic. 

\subsection{Concentration around the classical mean-field SIR model}
\label{subsec:concentration_sim}

In our first set of simulations, we validate Theorem \ref{thm:rapidly_mixing}, which says that if the interaction matrix $W$ has a spectral gap close to 1, $y_{av}^N(t)$ (and therefore $Y_{av}^N(t)$ in light of Theorem \ref{thm:concentration}) will concentrate around the classical mean-field SIR model. We generate interaction matrices derived from Erd\H{o}s-R\'{e}nyi random graphs for two reasons: they have a known expander-type property which leads to \eqref{eq:rapidly_mixing} being satisfied (see Example \ref{ex:ER}) and it is a commonly-used model for real world networks. 

\begin{figure}
\centering
\includegraphics[width=0.35\textwidth]{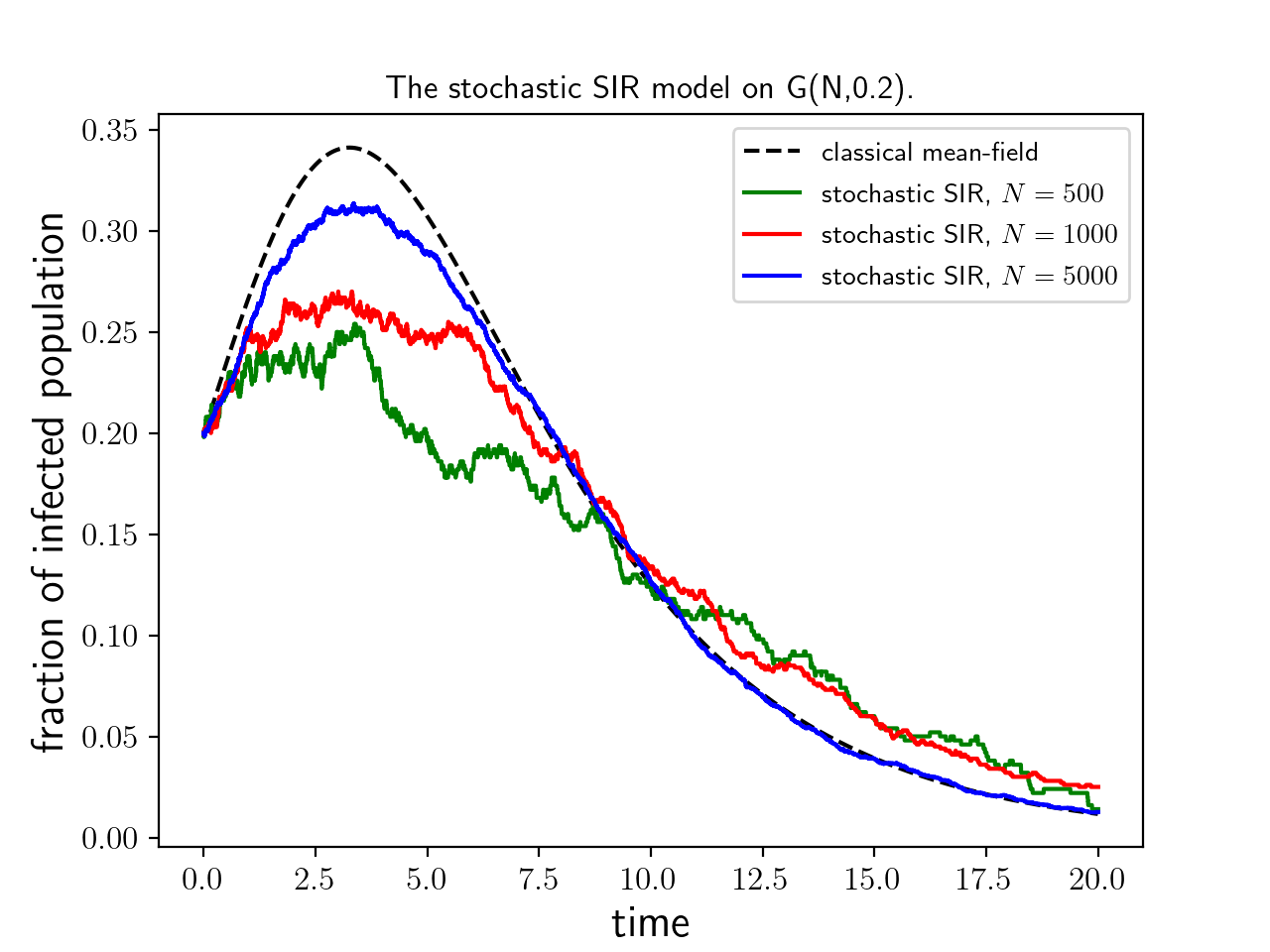}
\caption{Comparison of the classical SIR model with simulations of $Y_{av}^N(t)$ corresponding to the stochastic SIR process on realizations of $G(N, 0.2)$, for various values of $N$. }
\label{fig:erdos_renyi}
\end{figure}

Our simulations are set up as follows. We initially chose $0.2 \ast N$ vertices arbitrarily to be initially infected. Next, we sample the underlying contact network from $G(N,0.2)$, which denotes an Erd\H{o}s-R\'{e}nyi random graph on $N$ vertices with connection probability 0.2. Letting $d_i$ and $d_{max}$ be the maximum degree and the degree of $i$ in the contact network, respectively, the entries of the interaction matrix $W$ are given by 
$$
w_{ij} := \begin{cases}
\frac{1}{d_{max} + 1} & (i,j) \text{ is an edge in $G(N,0.2)$};\\
1 - \frac{d_i}{d_{max} + 1} & i = j \\
0 & \text{ else}.
\end{cases}
$$
It can be verified that $W$ is doubly stochastic and $\lambda(W) \le O ( \sqrt{\log N / N} )$ with high probability when $N$ is sufficiently large \cite[Appendix E]{sridhar2021meanfield}. In Figure \ref{fig:erdos_renyi}, we compare plots of the classical SIR model and the stochastic SIR process for $N = 500, 1000, 5000$. As predicted by Theorem \ref{thm:rapidly_mixing}, the stochastic processes converge to the classical SIR model when $N$ is sufficiently large. 

\begin{figure}
\centering
\includegraphics[width=0.35\textwidth]{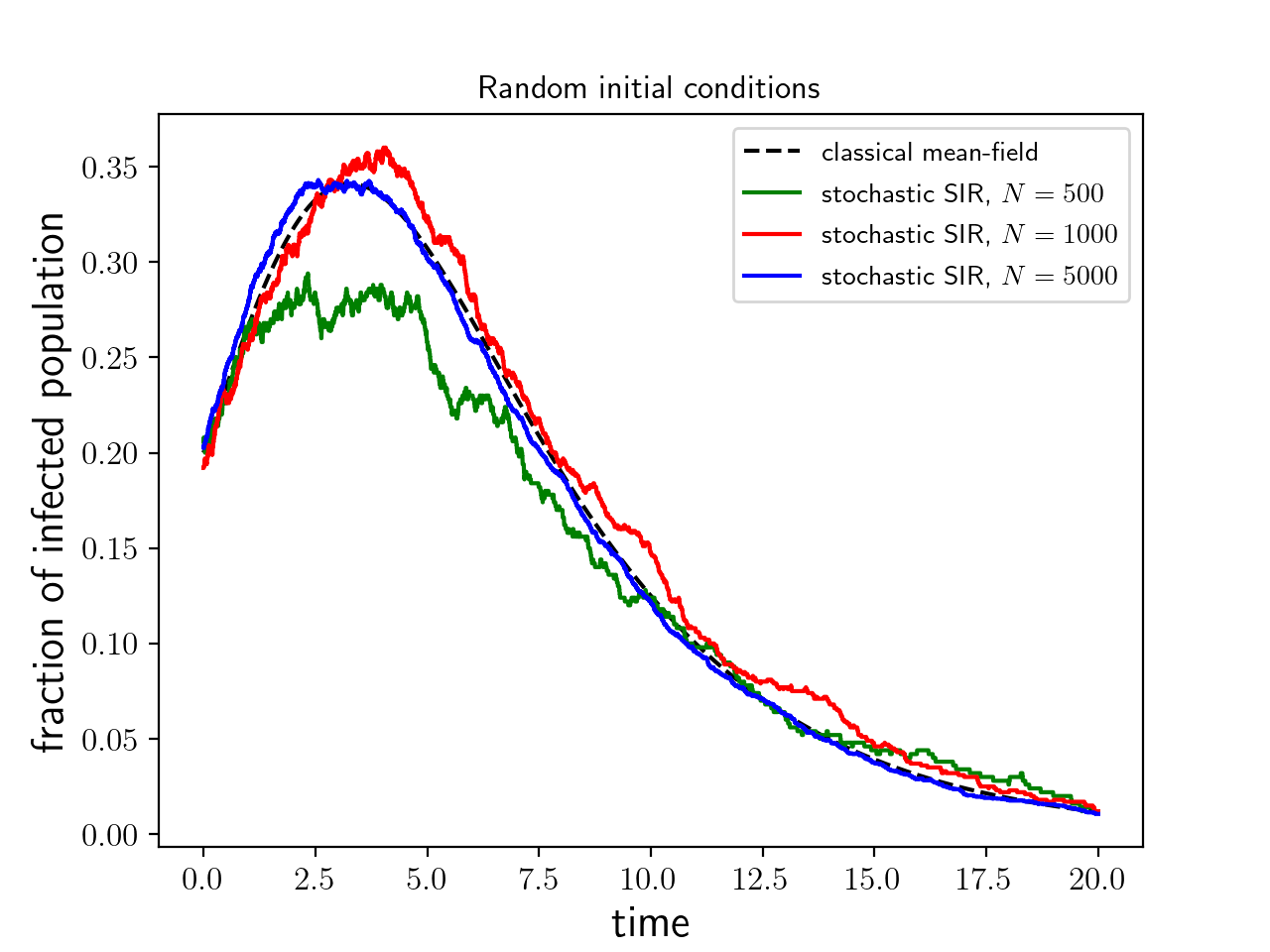}
\caption{Comparison of the classical SIR model with simulations of $Y_{av}^N(t)$ corresponding to the stochastic SIR process on nearest-neighbor networks with random initial conditions.}
\label{fig:nn_random}
\end{figure}

Next, we demonstrate that for general $W$ for which $\lambda(W)$ is {\it not} close to 0, the population state $Y_{av}^N(t)$ is close to the mean-field process if the initial set of infections is randomly chosen. The contact network, with parameters $(N,d)$ is constructed as follows: place the $N$ vertices at equidistant locations on the unit circle, and each vertex is connected to the $d$ closest vertices, including itself (we may assume $d$ is odd so that this construction is well-defined). For the interaction matrix, we set $w_{ij} = 1/d$ if $(i,j)$ is an edge in the contact network, else $w_{ij} = 0$. In \cite[Appendix E]{sridhar2021meanfield}, it was shown that if $\lim_{N \to \infty} d/N \in (0,1)$, then $\lambda(W)$ is bounded away from 0 even in the limit of large networks. Hence such networks do not fall under the purview of Theorem \ref{thm:rapidly_mixing}. The set of initial set of infections is chosen randomly in the manner of Example \ref{ex:random_initial_conditions}: each agent is infected with probability 0.2, independently across all agents. In Figure \ref{fig:nn_random}, we compare plots of the classical SIR model and the stochastic SIR process for $N = 500, 1000, 5000$ where $d = 0.2\ast N$. We see that the stochastic processes enjoy better concentration around the classical SIR model which validates Theorem \ref{thm:consensus_stable}.

\subsection{Deviation from the classical model}
\label{subsec:deviation_sim}

\begin{figure}
\centering
\includegraphics[width=0.35\textwidth]{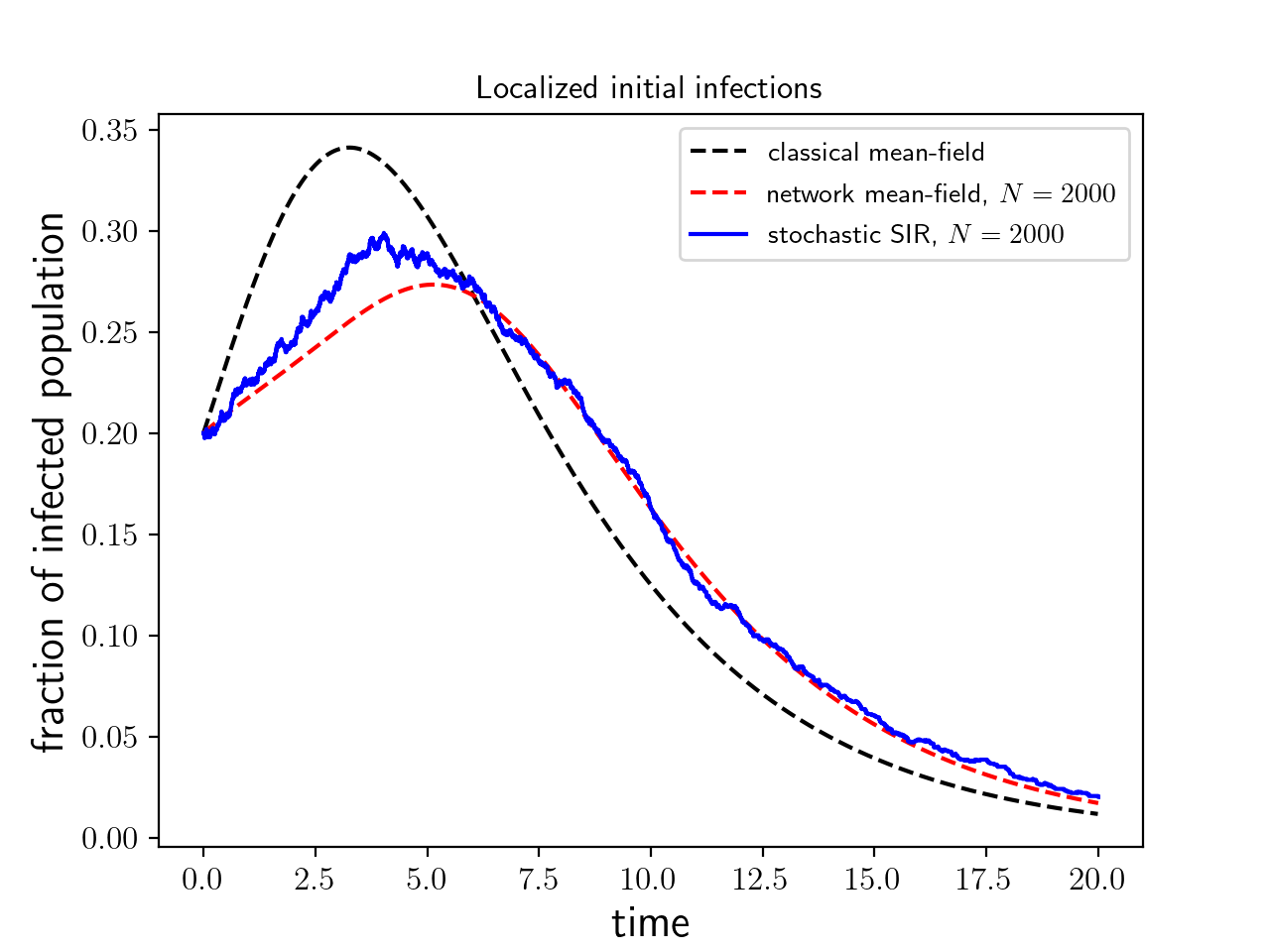}
\caption{Comparison of the classical SIR model, network SIR model and $Y_{av}^N(t)$ corresponding to the stochastic SIR process on a nearest neighbor graph with $N = 2000$ and $d = 400$.}
\label{fig:nn_localized}
\end{figure}

When $\lambda(W)$ is bounded away from 0 and the initial locations of infections are {\it not} well interspersed in the population, the evolution of $y_{av}^N(t)$ and therefore $Y_{av}^N(t)$ may be quite different from the classical SIR model. To illustrate this, we consider an interaction matrix constructed from the same nearest-neighbor graph family simulated in Figure \ref{fig:nn_random}, with parameters $N = 2000$ and $d = 0.2\ast N = 400$. We choose $400$ consecutive vertices on the unit circle to be the initial infected population.\footnote{By symmetry of the contact network, it does not matter which 400 vertices are chosen as long as they are consecutive.} In Figure \ref{fig:nn_localized}, we plot the classical SIR model, the network SIR model, and the stochastic SIR process corresponding to the interaction matrix we have defined. We see that the two mean-field approximations yield different predictions, with $Y_{av}^N(t)$ concentrating around the network SIR model rather than the classical model. Interestingly, $y_{av}^N(t)$ has a delayed, shorter infection peak compared to the prediction of the classical SIR model. Moreover, the networks considered in Figures \ref{fig:nn_localized} and \ref{fig:erdos_renyi} have approximately the same number of edges, illustrating how the behavior of $Y_{av}^N(t)$ is heavily influenced by fundamental structural properties of the network.

\begin{figure}
\centering
\includegraphics[width=0.35\textwidth]{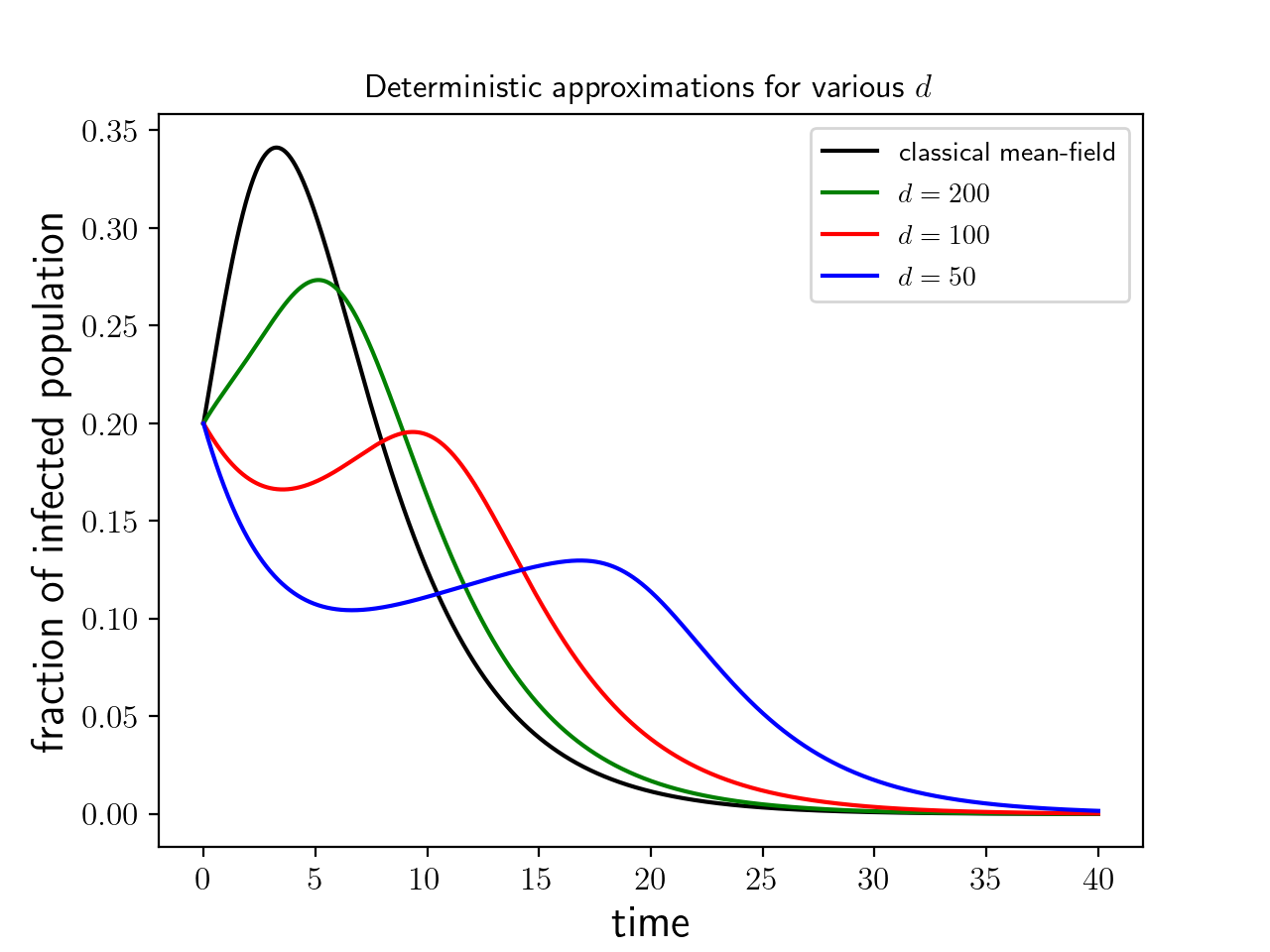}
\caption{Comparison of the classical SIR model with the network SIR model on nearest neighbor graphs with $N = 1000$ and varying $d$.}
\label{fig:nn_deterministic}
\end{figure}

Finally, we show how the network SIR model can exhibit quite different behaviors than what is predicted by the classical SIR model. As in the previous simulations, we let $N = 1000$ and choose 200 consecutive vertices to be the initial infected population. In Figure \ref{fig:nn_deterministic}, we compare the time-evolution of the classical SIR model as well as the network SIR model with $d = 200, 100, 50$. In the classical SIR model it is known that if an epidemic emerges, there is a peak in infections followed by an exponential decay to the zero-infection state \cite[Lemma 6]{mei_network_epidemics}. This behavior is exhibited when $d = 200$, though with a smaller, delayed peak in infections. Interestingly, when $d = 50, 100$, there is an initial {\it decline} in the number of infections followed by a resurgence in infections at a later point in time before experiencing an exponential decay to the zero-infection state. This behavior can be explained as follows. Since the initial infections are consecutive vertices and the graph is relatively sparse, many infected vertices will only have infected neighbors; hence these vertices cannot contribute to the epidemic and eventually die out; this explains the initial decline in the infection curve. Concurrently, the vertices that {\it do} have susceptible neighbors will spread the virus, leading to an increase in infections in other parts of the network; these two effects balance each other out at some point, leading to the observed valley in the infection curves. Surprisingly, the emergence of ``infection waves" observed in the {\it population-level} behavior of the $d = 50, 100$ plots has not been previously studied in network SIR models, theoretically or empirically.

\section{Conclusion}
\label{sec:conclusion}

In this paper, we investigated the connections between the stochastic SIR process, the classical SIR model and the network SIR model from a mathematically rigorous point of view. We showed that in general, the network SIR model provides a better approximation for the stochastic SIR process than the classical model, but that the classical model still yields correct predictions when the underlying network has an expander-type property or when the initial infections are well-mixed within the population. We also validated our results through simulations and empirically highlighted significant differences in the spreading behavior in the network SIR model compared to the classical SIR model. There are many avenues for future work, including a characterization of $y_{av}^N(t)$ for various types of networks as well as a theoretical study of the phenomena highlighted in our simulations.

\bibliographystyle{./bibliography/IEEEtran}
\bibliography{citations}

\end{document}